\documentclass[amsmath,amssymb,pra,aps,showpacs,superscriptaddress,twocolumn]{revtex4-1}
\pdfoutput=1
\usepackage{amsmath,amsfonts,amssymb,amsthm,graphics,graphicx,epsfig,bbm}
\usepackage[colorlinks=true,citecolor=blue,linkcolor=blue,urlcolor=blue]{hyperref}

\usepackage[usenames]{color}
\usepackage{graphicx}
\usepackage{subfigure}
\usepackage{amsmath}
\usepackage{epsfig}
\usepackage{dcolumn}
\usepackage{bm}
\usepackage{color}
\usepackage{epstopdf}
\usepackage{amssymb}
\usepackage{amstext}
\usepackage{latexsym}
\usepackage{hyperref}
\usepackage{amsfonts}
\usepackage{psfrag}
\usepackage{xcolor}
\usepackage[normalem]{ulem}
\usepackage{dsfont}
\usepackage{txfonts}
\usepackage{cases}
\usepackage{ifthen}

\graphicspath{{./Images/},{./ImagesAppendix/}}

\newcommand{\ket}[1]{\ensuremath{\left\vert #1 \right\rangle}}
\newcommand{\bra}[1]{\ensuremath{\left\langle #1 \right\vert}}

\newcommand{\braket}[2]{\langle #1 \vert #2 \rangle}

\newcommand{\an}[2]{\ensuremath{\hat{#1}^{\protect\phantom{\dagger}}_{#2}}}
\newcommand{\cn}[2]{\ensuremath{\hat{#1}^\dagger_{#2}}}
\newcommand{\nn}[2]{\ensuremath{\hat{n}^{#1}_{#2}}}

\newcommand{\expU}[1]{\ensuremath{e^{#1}}}
\newcommand{\abs}[1]{\left|#1\right|}

\newcommand{\pdif}[2]{\ensuremath{\frac{\partial#1}{\partial#2}}}
\newcommand{\avg}[1]{\ensuremath{\langle#1\rangle}}

\newcommand{\op}[1]{\ensuremath{\hat{#1}}}

\newcommand{\sectionappendix}[1]{\section{#1}}

\newcommand{\rev}[1]{{#1}}
\newcommand{\revB}[1]{#1}
\newcommand{\revC}[1]{#1}
\newcommand{\revD}[1]{#1}
\newcommand{\revE}[1]{#1}
\newcommand{\subfigimg}[3][,]{%
	\setbox1=\hbox{\includegraphics[#1]{#3}}
	\leavevmode\rlap{\usebox1}
	\rlap{\hspace*{2pt}\raisebox{\dimexpr\ht1-0.5\baselineskip}{{\bfseries \large\textsf{#2}}}}
	\phantom{\usebox1}
}

\newcommand\numberthis{\addtocounter{equation}{1}\tag{\theequation}}

\newcommand{\idg}[1]{{\bfseries #1)}}

\begin{document}
	
\title{The Aharonov-Bohm effect in mesoscopic Bose-Einstein condensates}

\author{Tobias Haug}
\affiliation{Centre for Quantum Technologies, National University of Singapore,
	3 Science Drive 2, Singapore 117543, Singapore}
\author{Hermanni Heimonen}
\affiliation{Centre for Quantum Technologies, National University of Singapore,
	3 Science Drive 2, Singapore 117543, Singapore}
\author{Rainer Dumke}
\affiliation{Centre for Quantum Technologies, National University of Singapore, 3 Science Drive 2, Singapore 117543, Singapore}
\affiliation{Division of Physics and Applied Physics, Nanyang Technological University, 21 Nanyang Link, Singapore 637371, Singapore}
\affiliation{MajuLab, CNRS-UNS-NUS-NTU International Joint Research Unit, UMI 3654, Singapore}
\author{Leong-Chuan Kwek}
\affiliation{Centre for Quantum Technologies, National University of Singapore,
	3 Science Drive 2, Singapore 117543, Singapore}
\affiliation{MajuLab, CNRS-UNS-NUS-NTU International Joint Research Unit, UMI 3654, Singapore}
\affiliation{Institute of Advanced Studies, Nanyang Technological University,
	60 Nanyang View, Singapore 639673, Singapore}
\affiliation{National Institute of Education, Nanyang Technological University,
	1 Nanyang Walk, Singapore 637616, Singapore}

\author{Luigi Amico}
\affiliation{Centre for Quantum Technologies, National University of Singapore,
	3 Science Drive 2, Singapore 117543, Singapore}
\affiliation{MajuLab, CNRS-UNS-NUS-NTU International Joint Research Unit, UMI 3654, Singapore}
\affiliation{Dipartimento di Fisica e Astronomia, Via S. Sofia 64, 95127 Catania, Italy}
\affiliation{CNR-MATIS-IMM \&   INFN-Sezione di Catania, Via S. Sofia 64, 95127 Catania, Italy}
\affiliation{LANEF {\it 'Chaire d'excellence'}, Universit\`e Grenoble-Alpes \& CNRS, F-38000 Grenoble, France}



\date{\today}

\begin{abstract}

\revC{Ultra-cold  atoms in light-shaped potentials open up new ways to explore mesoscopic physics: Arbitrary trapping potentials can be engineered with only a change of the laser field.  Here, we propose using ultracold atoms in light-shaped potentials to feasibly realize a cold atom device} to study one of the fundamental problems of mesoscopic physics, the Aharonov-Bohm effect: The interaction of particles  with a magnetic field when traveling in a closed loop. 
Surprisingly, we find that the Aharonov-Bohm effect is washed out for interacting bosons, while it is present for fermions.
We show that our atomic device has possible applications as quantum simulator, Mach-Zehnder interferometer and for tests of quantum foundation.
\end{abstract}


\maketitle
The Aharonov-Bohm effect is one of the most striking manifestations of quantum mechanics: Due to  phase shifts in the wave  function, specific interference effects arise when charged particles  enclose a region with a  non vanishing magnetic field\cite{Aharonov-Bohm}.
\rev{This effect has important implications}
in  foundational aspects of quantum physics\cite{Aharonov-Bohm,AB_fundations_RMP,AB_fundations_Vaidman,leggett1980macroscopic} and many-body quantum physics\cite{lobos2008effects, 
rincon2008spin
,shmakov2013aharonov,hod2006inelastic,jagla1993electron}. The Aharonov-Bohm effect has been influential 
\rev{in many fields}
of physical sciences, like mesoscopic physics, quantum electronics and molecular electronics\cite{gefen1984quantum,buttiker1984quantum,webb1985observation,nitzan2003electron}, with remarkable \rev{applications
enabling}
quantum technologies\cite{byers1961theoretical,bloch1968flux,gunther1969flux,bachtold1999aharonov,coskun2004h,cardamone2006controlling}. 

An electronic fluid confined to a ring-shaped wire pierced by a magnetic flux is  the typical configuration employed  to study the Aharonov-Bohm effect. In this way, a matter-wave interferometer is realized:
The current through the ring-shaped quantum system displays characteristic oscillations depending on the imparted magnetic flux.  Neutral particles with magnetic moments display similar interference effects\cite{AharonovCasher}. 


A new perspective to study the transport through  small and medium sized quantum matter systems \rev{has been demonstrated recently in }
ultracold atoms\cite{papoular2014fast,li2016superfluid,
krinner2015observation,husmann2015connecting}: 
\rev{In such systems, it is possible for the first time to manipulate and adjust 
the carrier statistics, particle-particle interactions and  spatial configuration of the circuit. 
\revC{Such flexibility is very hard, if not impossible, to achieve using  standard realizations of mesoscopic systems. Mesoscopic phenomena are studied predominantly with electrons in condensed matter devices. The range of parameters that can be explored is limited since a single change in a parameter requires a new device or may not be possible at all. 
To adjust all those parameters } Atomtronics has been put forward\cite{seaman2007atomtronics,Amico_Atomtronics,Amico_focus}}.

\rev{In this paper, we study the Aharonov-Bohm  effect in a mesoscopic ring-shaped bosonic condensate pierced by a synthetic magnetic flux\cite{dalibard2011colloquium}: The bosonic fluid is injected  from  a `source' lead, propagates along the ring,  and it is collected in a `drain' lead.   In this way, we provide the atomtronic counterpart of an iconic problem in mesoscopic physics\cite{gefen1984quantum,buttiker1984quantum}, with  far reaching implications over the years in the  broad area of physical science\cite{lobos2008effects, 
rincon2008spin,
shmakov2013aharonov,hod2006inelastic,jagla1993electron,byers1961theoretical,bloch1968flux,gunther1969flux,bachtold1999aharonov,coskun2004h,cardamone2006controlling}.  
%
This system can realize an elementary component of an atomtronic integrated circuit\cite{Boshier_integrated}}.
\rev{We analyse the non-equilibrium dynamics of the system by  quenching the particles spatial confinement; our study is combined with the analysis of the out-of-equilibrium dynamics  triggered by driving the current through suitable baths attached to the system \revB{within Markovian approximations and an exact simulation using DMRG}.}   
Depending  on the ring-lead coupling, interactions  and  particle statistics,  the  system displays qualitatively distinct non-equilibrium regimes characterized by different response of the interference pattern  to the effective gauge field.  
Remarkably, the interacting bosonic system lacks the fundamental Aharonov-Bohm effect as it is washed out, in contrast to a fermionic system.
\revB{Finally, we explore possible applications of this device to realize new atomtronic quantum devices, quantum simulators and tests for quantum foundation.} 
\section{Model}
\begin{figure}[htbp]
	\centering
	\includegraphics[width=0.49\textwidth]{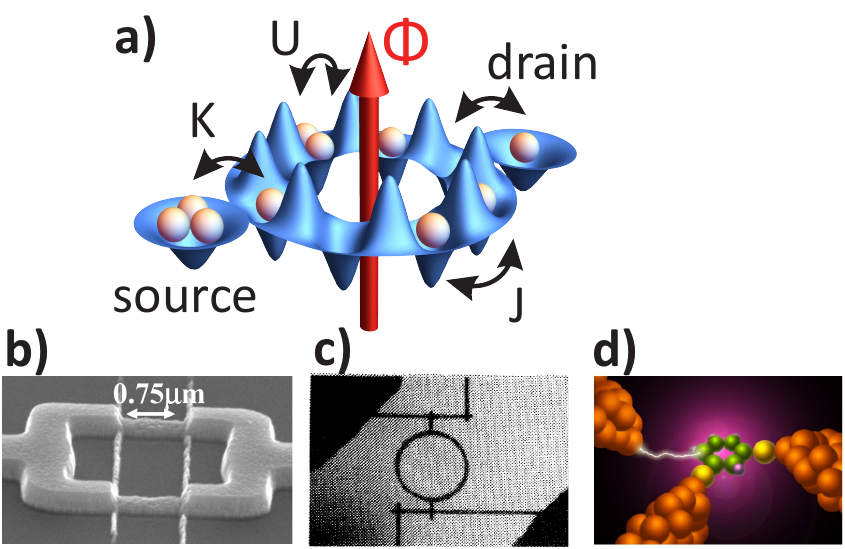}
	\caption{{\bfseries Mesoscopic systems and its analogous atomtronic architecture} \idg{a} Atomtronic setup consisting of a superfluid condensate in a ring lattice with two attached leads. The dynamics is controlled by Aharonov-Bohm flux $\Phi$ and ring-lead coupling $K$. Atoms tunnel between ring sites with rate $J$ and interact on-site with strength $U$.
	Related mesoscopic condensed matter devices to study Aharonov-Bohm effect are \idg{b} superconducting interference devices\cite{angers2008proximity} \idg{c} nanoscopic metal rings \cite{webb1985observation} \idg{d} proposed molecular quantum device \cite{cardamone2006controlling}.}
	\label{Model}
\end{figure}
The Bose-Hubbard Hamiltonian ${\mathcal{H}=\mathcal{H}_\text{r}+\mathcal{H}_\text{l}}$ describes the system consisting of a ring with an even number of lattice sites $L$ and two leads (see Fig.\ref{Model}).
The ring Hamiltonian is given by
\begin{equation}
\mathcal{H}_\text{r}=-\sum_{j=0}^{L-1}\left(J\expU{i2\pi\Phi/L}\cn{a}{j}\an{a}{j+1} + \text{H.C.}\right)+\frac{U}{2}\sum_{j=0}^{L-1}\nn{}{j}(\nn{}{j}-1)\;,
\end{equation}
where $\an{a}{j}$ and $\cn{a}{j}$ are the annihilation and creation operator at site $j$, $\nn{}{j}=\cn{a}{j}\an{a}{j}$ is the particle number operator, $J$ is the intra-ring hopping, $U$ is the on-site interaction between particles and $\Phi$ is the total flux through the ring. Periodic boundary conditions are applied: $\cn{a}{L}=\cn{a}{0}$.

The two leads dubbed source (S) and drain (D) consist of a single site each, which are coupled symmetrically at opposite sites to the ring with coupling strength $K$. \rev{In  both of them, local potential energy and  on-site interaction are set to zero as the leads are considered to be large with low atom density.} The lead Hamiltonian is ${\mathcal{H}_\text{l}=-K(\cn{a}{S}\an{a}{0} +\cn{a}{D}\an{a}{L/2} +\text{H.C.})}$,
where $\cn{a}{\text{S}}$ and $\cn{a}{\text{D}}$ are the creation operators of source and drain respectively.


The system is initially prepared with all particles in the source and the dynamics is strongly affected by the lead-ring coupling.
We calculate the state at time $t$ with ${\ket{\Psi(t)}=\expU{-i\mathcal{H}t}\ket{\Psi(0)}}$. We investigate the expectation value of the density in source and drain over time, which for the source is calculated as ${n_\text{source}(t)=\bra{\Psi(t)}\cn{a}{S}\an{a}{S}\ket{\Psi(t)}}$ and similar for the drain.
\rev{We point out that, by construction, our approach is well defined  for the whole cross-over ranging  from the weak to strong leads-system coupling (in contrast with the limitations of traditional approaches \rev{for interacting particles} mostly valid for  the regime of weak lead-system coupling\cite{tokuno2008dynamics}). \revE{We assume that the motion of the atoms involves only the lowest Bloch-band, thus providing a purely one-dimensional dynamics. Our results are given in units of the tunneling rate $J$ between neighboring ring sites. It depends exponentially on the lattice spacing. In state-of-the-art experiments on cold atoms in lattices, ${J/\hbar\approx 250-500\text{Hz}}$ was reported\cite{atala2014observation,aidelsburger2015measuring} and atom lifetimes of 8s\cite{aidelsburger2015artificial}. In experiments, this would restrict the maximal observation time $t$ in units of $J$ to ${tJ=2000-4000}$. }}

\section{Results}
In the weak-coupling regime ${K/J\ll1}$, the lead-ring tunneling is slow compared to the dynamics inside the ring. In this regime, the condensate mostly populates the drain and source, leaving the ring nearly empty. As a result, the scattering due to on-site interaction $U$ has a negligible influence on the dynamics.  
\begin{figure}[htbp]
	\centering
	\includegraphics[width=0.49\textwidth]{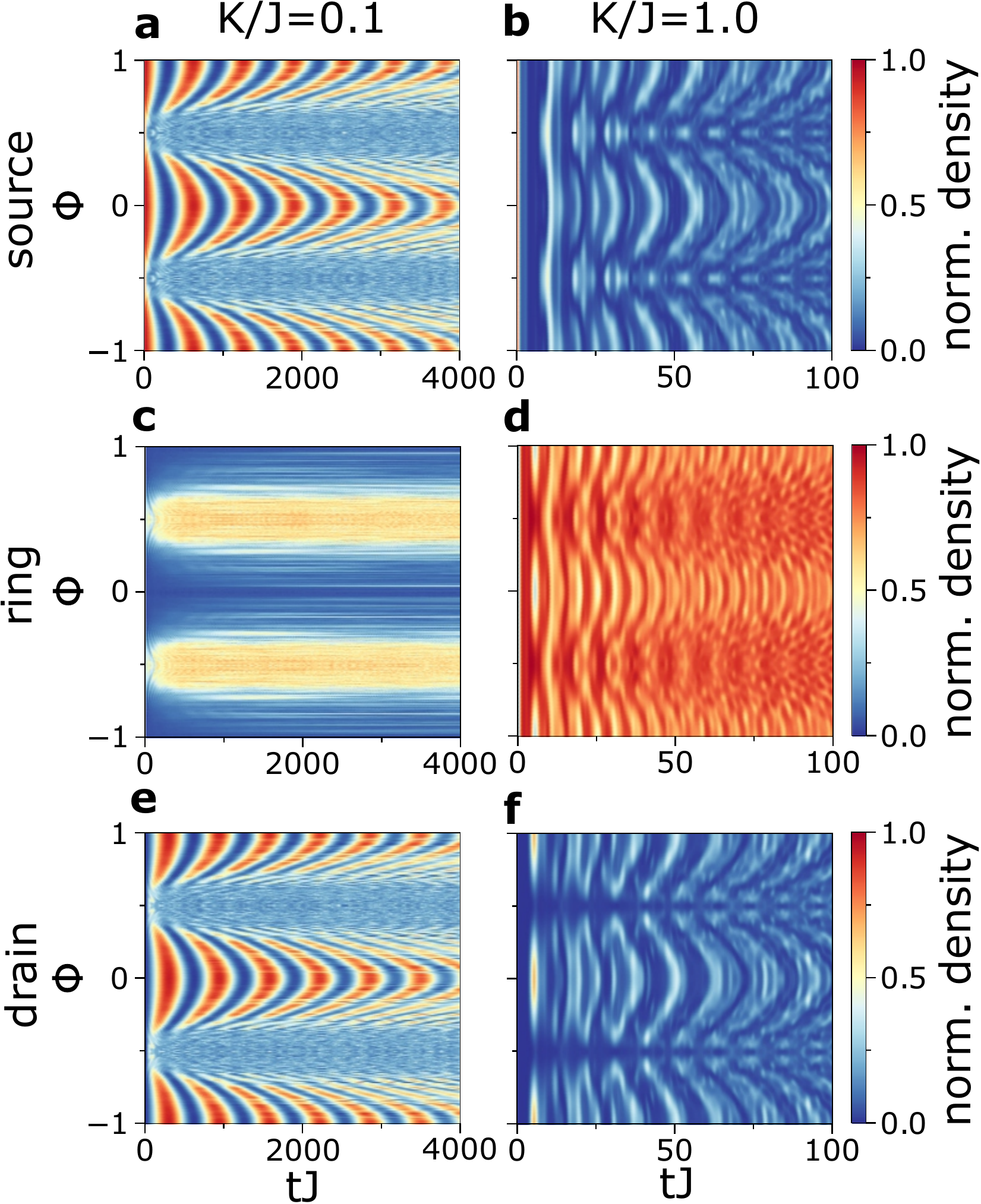}
	\caption{{\bfseries Time evolution of density} in source \idg{a,b}, ring \idg{c,d} and drain \idg{e,f} plotted against flux $\Phi$. \idg{a,c,e} weak ring-lead coupling ${K/J=0.1}$ (on-site interaction ${U/J=5}$). \idg{b,d,f} strong ring-lead coupling ${K/J=1}$ (${U/J=0.2}$). Time is indicated $tJ$ in units of inter-ring tunneling parameter $J$. The number of ring sites is ${L=14}$ with $N_\text{p}=4$ particles initially in the source. The density in the ring is ${n_\text{ring}=1-n_\text{source}-n_\text{drain}}$.}
	\label{CentralFigure}
\end{figure}
With increasing $\Phi$ the oscillation becomes faster and the ring populates, resulting in increased scattering and washed-out density oscillations.

In the strong-coupling regime ${K/J\approx1}$, the  lead-ring  and the intra-ring dynamics are characterized by  the same frequency  and cannot be treated separately. \rev{Here, a superposition of many oscillation frequencies appears (see also supplementary material), and after a short time} the condensate is evenly spread both in leads and ring (Fig.\ref{CentralFigure}d,e,f). The density in the ring is large and scattering affects the dynamics by washing out the oscillations. Close to ${\Phi=0.5}$, the oscillations slow down, especially for weak interaction, due to destructive interference\cite{valiente2008two}. \revE{We studied the dynamics of the relative phase between source and drain: We find that relative phase displays similar dynamics as source and drain density (see supplemental materials).}

\begin{figure}[htbp]
	\centering
	\subfigure{\includegraphics[width=0.46\textwidth]{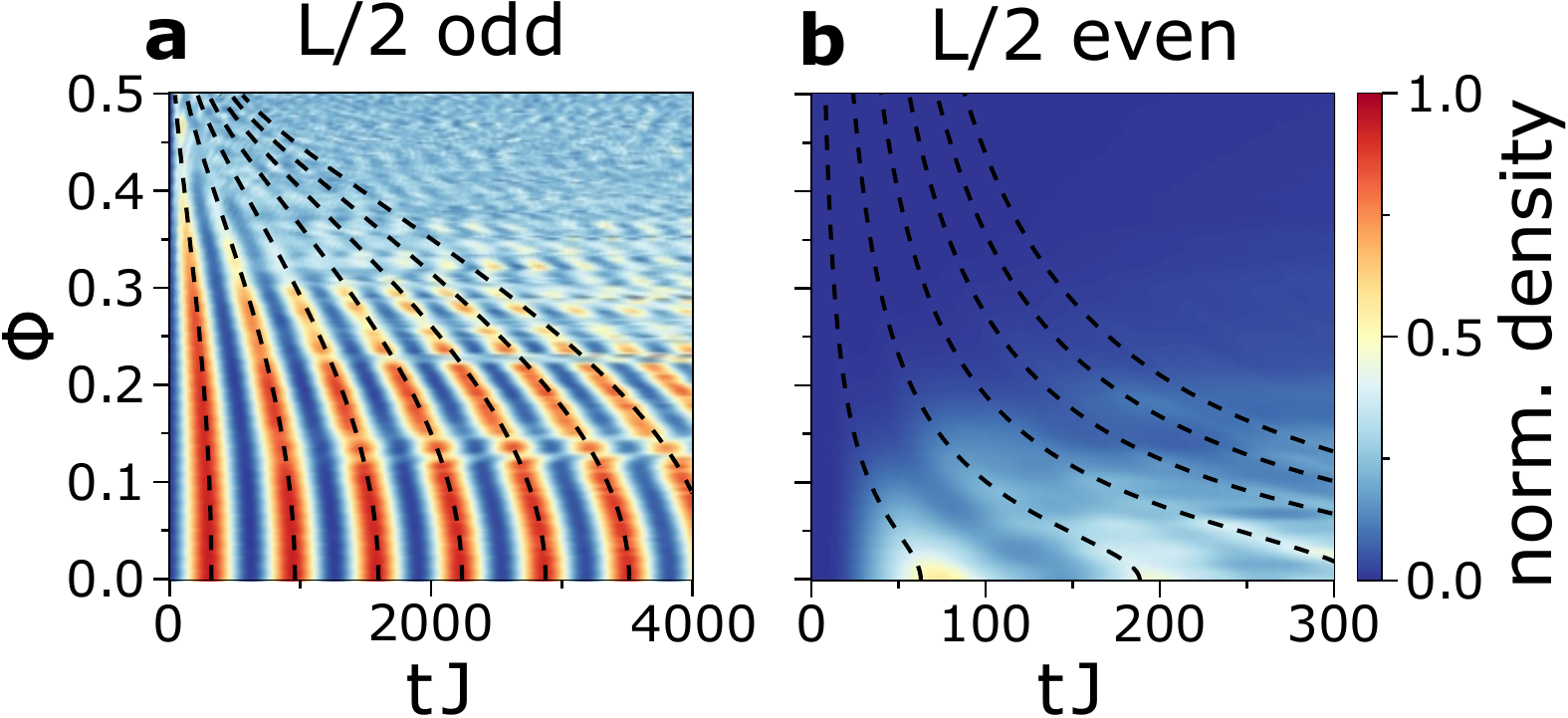}}
	\caption{{\bfseries Density in drain against flux $\Phi$ and parity in number of ring sites} odd parity \idg{a} ${L/2=7}$ (${U/J=3}$) and even parity \idg{b} ${L/2=8}$ (${U/J=1}$). Simulation with ${N_\text{p}=4}$ and weak coupling (${K/J=0.1}$). The structures around ${\Phi=0.15}$ for odd parity are many-body resonances\cite{sturm2017quantum}.  Dashed line shows analytic derived oscillation  period ({\bfseries a} Eq.\ref{FreqFull4n2Higher} and {\bfseries b} Eq.\ref{Freq4nApp} in the supplementary material).
}
	\label{ControlParity}
\end{figure}

We also find that the dynamics is affected by the parity in half of the number of ring sites $L/2$ especially in the weak-coupling regime.
In Fig.\ref{ControlParity}, we find that for odd (${L/2=3,5,7,\dots}$) and even parity (${L/2=2,4,6,\dots}$) the flux dependence and time scales differ widely.  
\rev{Similar to tunneling through quantum dots, we can understand the parity effect in terms of ring-lead resonant and off-resonant coupling\cite{glazman2003coulomb}.
Off-resonant coupling is characterized by regular, slow oscillations between source and drain and a small ring population. 
Resonant coupling implies faster oscillation, but a large ring population. The resulting dynamics is affected  by the interplay between interaction $U$ and $\Phi$. The flux $\Phi$ modifies the energy eigenmodes of the ring, bringing them in and out of resonance with the leads. Interaction $U$ washes out the oscillations between source and drain when the ring population is large.
For odd parity, we find that both resonant and off-resonant coupling contributes. Close to ${\Phi=0}$ the  off-resonant coupling dominates and due to the small ring population interaction 
has only a minor effect on the dynamics.  
Close to ${\Phi=0.5}$, resonant ring modes become dominant, and the faster oscillations are washed out by the higher ring population.
For even parity only resonant coupling is possible. Close to ${\Phi=0}$, ring modes are on resonance, resulting in fast oscillations washed out by interaction. For increasing $\Phi$ transfer is suppressed as ring modes move out of resonance and off-resonant coupling is not possible (detailed derivation in supplementary materials). }
Parity effects are suppressed with strong coupling or many ring sites as the level spacing decreases and many ring modes can become resonant. 

\begin{figure}[htbp]
	\centering
	\subfigure{\includegraphics[width=0.49\textwidth]{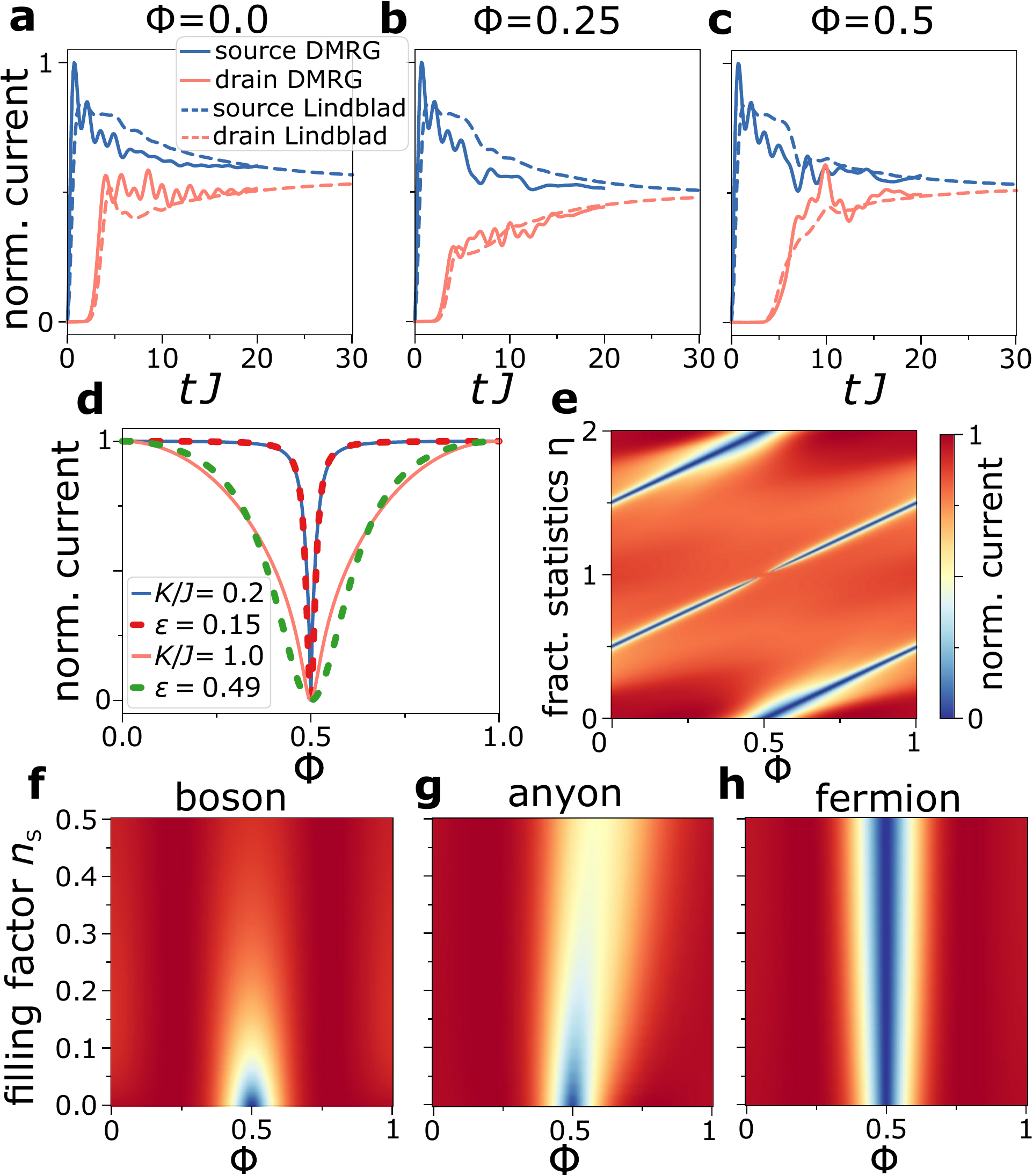}}
	\caption{{\bfseries Current through the Aharonov-Bohm ring} \revB{\idg{a-c} Evolution of source and drain current towards the steady state (when both currents are the same) with DMRG (solid line) and Lindblad formalism (dashed) for hard-core bosons, ${K=1}$ and ${L_\text{R}=10}$. For DMRG, both reservoirs and ring are solved with Schr\"odinger equation as a closed system. Source and drain are modeled as chains of hard-core bosons with equal length ${L_\text{S}=L_\text{D}=30}$. \revC{Initially, the source is prepared at half-filling ($N_\text{p}=15$) in its ground state (ring and drain are empty) decoupled from the ring (${K(t=0)=0}$). For ${t>0}$ the coupling is suddenly switched on (${K(t>0)=J}$).} Due to numerical limitations, we analyse the short-time dynamics. For the open system, the reservoirs obey Pauli-principle with ${r=0.65}$ and ${\Gamma=1.5}$.}
	\rev{\idg{d} Solid lines: steady-state current ($j_\text{SS}$) we obtained applying the method presented in \cite{prosen2008third,guo2017solutions} for non-interacting particles with ${L=100}$. Dashed lines: a fit (${\epsilon=\{0.15,0.49\}}$) with the transmission equations derived by B\"uttiker et al.\cite{buttiker1984quantum}. }
	\idg{e} $j_\text{SS}$ for infinite on-site interaction in both leads and ring  plotted against flux $\Phi$ and fractional statistics $\eta$ (${\eta=\{0,2\}}$ non-interacting fermions, ${\eta=1}$ hard-core bosons, else anyons) for strong source-drain imbalance. \rev{The reservoirs obey the Pauli principle with ${r=0}$, ${\Gamma=1/2}$}. The number of ring sites is ${L/2=3}$ and the ring-lead coupling is ${K/J=1}$. 
	At the transition to bosons, there is a discontinuity in the current.
	\idg{f-h} $j_\text{SS}$ for hard-core bosons, anyons (${\eta=0.25}$) and fermions plotted against flux and the filling factor $n_\text{S}$.  \rev{The reservoirs can have multiple particles per state and have a small particle number imbalance  between source and drain with ${n_\text{S}-n_\text{D}=0.01}$. The current is normalized to one for each value of filling independently.}
    } 
	\label{currentStatGraph}
\end{figure}

{\em Open system--} To study the properties of a filled ring, 
in Fig.\ref{currentStatGraph} we couple particle reservoirs with the leads to drive a current through the now open system. \rev{We model it using the Lindblad master equation 
\begin{equation*}
	\pdif{\rho}{t}=-\frac{i}{\hbar}\left[H,\rho\right]-\frac{1}{2}\sum_m\left\{\cn{L}{m}\an{L}{m},\rho\right\}+\sum_m\an{L}{m}\rho\cn{L}{m}
\end{equation*}
 for the reduced density matrix (tracing out the baths) \cite{breuer2002theory}. The bath-lead coupling is assumed to be weak and within the Born-Markov approximation. We consider two types of reservoirs: The first type allows multiple particles per reservoir state ${L_1=\sqrt{\Gamma n_\text{S}}\cn{a}{S}}$, ${L_2=\sqrt{\Gamma (n_\text{S}+1)}\an{a}{S}}$, ${L_3=\sqrt{\Gamma n_\text{D}}\cn{a}{D}}$ and  ${L_4=\sqrt{\Gamma (n_\text{D}+1)}\an{a}{D}}$ ($n_\text{S}$ ($n_\text{D}$) is the density of the source (drain) site if uncoupled to the ring). The other type is restricted to a single particle per state (Pauli-principle)
${L_1=\sqrt{\Gamma}\cn{a}{S}}$, ${ L_2=\sqrt{r\Gamma}\an{a}{S}}$, and ${L_3=\sqrt{ \Gamma}\an{a}{D}}$ ($r$ characterizes the back-tunneling into the source reservoir). 
We solve the equations for the steady state of the density matrix ${\pdif{\rho_\text{{SS}}}{t}=0}$ numerically\cite{guo2017dissipatively}. The current operator is ${j=-iK(\cn{a}{\text{S}}\an{a}{0}-\cn{a}{0}\an{a}{\text{S}})}$ and its expectation value is ${\avg{j}=\text{Tr}(j\rho_\text{SS})}$. We generalize the particle statistics with the parameter $\eta$ (${\eta=\{0,2\}}$ fermions, ${\eta=1}$ bosons, else anyons) using the transformation ${\cn{a}{n}\rightarrow\cn{a}{n}\prod_{j=n+1}^{L}\expU{i\pi (1-\eta)\nn{}{j}}}$\cite{amico1998one,keilmann2011statistically,greschner2015anyon}.} 
\revB{In Fig.4 a) -c), we compare the open system Lindblad approach with a full simulation of both ring and reservoirs using DMRG (Density Matrix Renormalization group, \revC{details in the caption and supplementary materials})\cite{white2004real,itensor}. Both methods yield similar results, with the Lindblad approach smoothing out the oscillation found in DMRG. This shows that leads modeled as Markovian bath without memory is sufficient to describe the dynamics. 
Using both methods, we calculate the evolution towards the steady-state. Remarkably, for the current the {\it initial dynamics  depends on the flux}, showing the Aharonov-Bohm effect of the dynamics. However we find surprisingly, that {\it the steady-state reached after long times is nearly independent of flux.}}

\rev{For vanishing atom-atom interactions, the equilibrium scattering-based results of B\"uttiker et al.\cite{buttiker1984quantum} and the non-equilibrium steady state current yield similar result -- Fig.\ref{currentStatGraph}d). 
Next, we enforce the Pauli-principle (${U=\infty}$) in both leads and ring and vary the particle statistics and the average number of particles in the system (filling factor). Fermions are then non-interacting, while anyons and bosons interact more strongly with increasing filling.} 
\rev{Now, we use the open system method to characterize the steady-state current. We found that the type of particle and inter-particle interaction has a profound influence on the Aharonov-Bohm effect--Fig.\ref{currentStatGraph} e) - h). While non-interacting fermions or bosons react strongly to an applied flux, interacting bosons have only weak dependence on the flux. 
Fermions have zero current at the degeneracy point, while anyons have a specific point with minimal current, which depends on the reservoir properties. 
When the filling of atoms in the ring is increased, fermions show no change in the current. However, for anyons a shift of the Aharonov-Bohm minimum in flux is observed. The minimum weakens the closer the statistical factor is to the bosonic exchange factor.
For hard-core bosons, we find that the current becomes minimal at half-flux for low filling, however {\it vanishes with increasing filling.} The scattering between atoms increases with the filling factor, washing out the Aharonov-Bohm effect. }

\begin{figure}[htbp]
	\centering
	\subfigure{\includegraphics[width=0.49\textwidth]{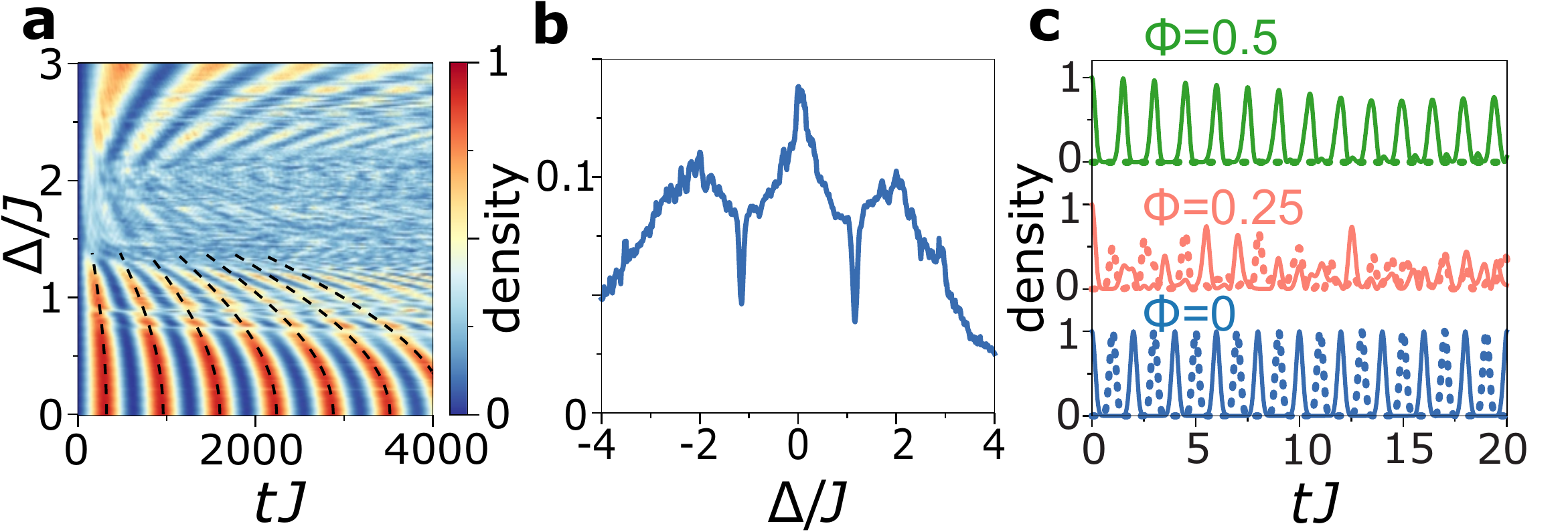}}
	\caption{{\bfseries Applications} \idg{a} Density in  drain plotted against two potential barriers placed symmetrically in both arms of the ring with depth $\Delta$. ${L=14}$, ${N_\text{p}=4}$, ${K/J=0.1}$ and ${U/J=3}$. Dashed line indicates a fit with the analytic formula for the oscillation period used for the flux dependence in Fig.\ref{ControlParity}a,b (Supplementary material Eq.\ref{FreqFull4n2Higher}, replace $\Phi$ with $\Delta/(2J)$).
	\idg{b}	Average density in drain integrated over a time ${t=10000/J}$ for a potential well with depth $\Delta$ in one arm of the ring. Interference effects cause minima in transmission rate for certain values of $\Delta$. ${L=14}$, ${N_\text{p}=4}$, ${K/J=1}$ and ${U/J=0.1}$.
	\idg{c} Density in source (solid) and drain (dashed) with the perfect state transfer protocol with ${U/J=0}$ and ${L=14}$. Atoms oscillate between source and drain with period ${t=2}$. 
	}
	\label{PSTFlux}
\end{figure}
\section{Discussion}
The dynamics of atoms in the ring device can be controlled with the ring-lead coupling and flux. 
In general, interaction between atoms washes out the well-defined oscillations of current between source and drain.
However, the effect of the interaction depends specifically also on the geometry. For odd parity, the interaction between the atoms does not have significant influence on the dynamics. 
Using the flux, it is possible to switch the transmission through the device for even parity.

We find that the current through this device depends strongly on the particle statistics. Fermions behave fundamentally different from bosons. Fermions show a strong Aharonov-Bohm effect, which has been studied in mesoscopic devices. However, interacting bosons have not been realized in a mesoscopic device.
{\it Remarkably,  for interacting bosons in the strong coupling regime  the Aharonov-Bohm  effect is effectively suppressed.} \revD{Indeed, the Aharonov-Bohm effect results from a gauge field that breaks time-reversal symmetry and modifies the phase of particles traveling along the two paths of the ring. Interacting bosons can condense   with the emergence of a condensate phase. Our results indicate  that this condensate phase is able to cancel the phase shift induced by the Aharonov-Bohm effect and suppress it in interacting bosons. Surprisingly, we find that even in the non-equilibrium dynamics we studied the Aharonov-Bohm effect remains suppressed. Our study  of the transport of anyonic particles confirms that the statistical factors  can modify the interference: the anyon statistical factor is found able to both move the Aharonov-Bohm minimum and weaken the dependence of the interference on the applied flux.}  

In summary, the Aharonov-Bohm effect in the mesoscopic regime does experience  a non-trivial cross-over as a function of interaction, carrier statistics and the ring-lead coupling strength. Using cold atoms, this device would allow the first time to observe these effects for bosons.

Here we present possible applications using the physics discussed above.  \rev{We study them in the closed ring-lead configuration, with the atoms initially in the source.} \revD{These devices could be readily realized in cold atom experiments.}

{\bfseries dc-SQUID: }First, we study the atomtronic counterpart of the dc-SQUID: 
\rev{We change the local potential by $\Delta$ at two single sites in the ring symmetrically in the upper and lower half by adding the following part to the Hamiltonian: $\mathcal{H}_\text{imp}=\Delta(\nn{}{\lceil L/4\rceil}+\nn{}{\lceil3L/4\rceil})$.}  The \rev{time-evolution depending} on $\Delta$ is shown in Fig.\ref{PSTFlux}a,b. The potential barrier modifies the transfer rate to the drain in a quantitatively similar way as the Aharonov-Bohm flux. However, no destructive interference is observed. This indicates that the barrier influences the dynamics only by scattering incoming particles, but does not imprint a phase shift. However, by adjusting $\Delta$ we can control the source-drain transfer rate \rev{in a similar fashion as the flux.} \revD{This device would realize an  easily controllable atomtronic transistor.}

{\bfseries Quantum dot simulator: }\rev{Next, we study the propagation through a quantum dot like structure\cite{yeyati1995aharonov,
wu1998resonant}. Here, the local potential is changed by adding a potential well on one arm of the ring $\mathcal{H}_\text{qd}=\Delta\sum_{j=0}^{(L-6+\text{mod}(L,4))/2}\nn{}{j}$.} We found that distinct transmission minima are displayed (see Fig.\ref{PSTFlux}c). Such results indicate that the atoms acquire a phase difference while traveling through the ring. \rev{This device could realize a switch by changing $\Delta$ around the transmission minima, or alternatively a simulator for quantum dots.}

{\bfseries Perfect state transfer: }Finally, we investigate the Perfect State Transfer protocol, where particles move from source to drain and vice-versa without dispersion at a fixed rate\cite{dai2009engineering}.  \rev{The coupling parameters are
$J_n = \frac{\pi}{2}J\sqrt{s_n}\sqrt{n(L_0-n)}$,
where $n$ is numeration of the coupling from source to drain, ${L_0=L/2+3}$ the number of sites on the shortest path between source and drain, and $s_n$ secures the Kirchhoff's law. We set $s_n=1$ everywhere except at the two ring sites which are coupled directly to the leads: There, the coupling of those sites to the neighboring two ring sites is $s_n=1/2$}. The flux dependence of the time evolution of the density for ${U=0}$ is shown in Fig.\ref{PSTFlux}c . At ${\Phi=0}$ we observe that the density in source and drain oscillates at a constant rate with close to unit probability.  \rev{Depending on interaction and particle number, the fidelity of the transport remains at unity or decreases. We will study this interesting effect in a future publication.} In contrast to weak coupling, the particles move as a wave packet inside the ring. By tuning the flux, the drain density can be controlled and transmission to the drain becomes zero at the degeneracy point.
The setup with perfect state transfer could realize a switch or  atomtronic quantum interference transistors: By changing the flux, perfect transmission is changed into perfect reflection.
\revB{We note that our system can be  relevant for  Mach-Zehnder matter-wave interferometer  with enhanced flexibility and control (see \cite{berrada2013integrated,ji2003electronic,sturm2014all}).  The setup is a new tool to to test quantum foundation with an interaction-free measurement.} 
In particular, we propose to use the high control over the dynamics to create an atomic version of a Elitzur-Vaidman bomb tester, the hallmark example of interaction-free measurement\cite{elitzur1993quantum}. The system is prepared with a single particle, the flux set to the degeneracy point and a bomb, which is triggered when the particle is measured in one specific arm of the ring. \rev{Without the bomb, the Aharonov-Bohm effect prevents the particle from reaching the drain.} Only if there is a bomb and the particle has not triggered it, the particle reaches the drain \rev{with unit probability due to the perfect state transfer.} This setup has a 50\% chance to detect the bomb without detonating it, improving from the 33\% efficiency of the photonic implementation.

\section{Conclusion}
\rev{We studied the  non equilibrium  transmission through an Aharonov-Bohm mesoscopic ring.   By quenching the spatial confinement,  the dynamics is strongly affected by the leads-ring coupling, the parity of the ring sites, and  the interaction of the atoms. By combining our analysis with the study of  the non-equilibrium steady states in an open system, we find that the Aharonov-Bohm effect  is washed out for interacting bosons. Finally, we have analyzed the possible implications of our study to conceive new quantum atomtronic devices.

\revB{We believe our study will be instrumental to bridge cold-atom and mesoscopic physics and create a tool to explore new areas of research.
In particular, our approach effectively defines new directions in quantum transport: important chapters of the field, like full counting statistics and shot noise\cite{lovas2017full}, matter-wave interferometers, rotation sensors and non-Markovian dynamics\cite{chiuri2012linear} could be studied with the new twist  provided by the cold atoms quantum technology.
}
Most of the physics we studied here could be explored experimentally with the current know-how in quantum technology and cold atoms.\revC{In particular, flux in ring condensates \cite{wright2013driving} or clock transitions\cite{lai2019photovoltaic}, lattice rings \cite{amico2014superfluid} and quench dynamics in leads\cite{eckel2016contact} have been demonstrated with recent light-shaping techniques\cite{gauthier2016direct}. Atom dynamics can be measured via fluorescence or absorption imaging of the density or current\cite{mathew2015self}.}} Our results can be relevant in other contexts of quantum technology, beyond ultracold atoms\cite{roushan2017chiral}.


\begin{acknowledgments}
We thank A. Leggett for enlightening discussions. The Grenoble LANEF framework (ANR-10-LABX-51-01) is acknowledged for its support with mutualized infrastructure. We thank National Research Foundation Singapore and the Ministry of Education Singapore Academic Research Fund Tier 2 (Grant No. MOE2015-T2-1-101) for support.
\end{acknowledgments}
\bibliography{library}
\appendix

\sectionappendix{Numerical methods}
{\bfseries Exact diagonalization} The low-lying energy states and dynamics of small closed and open system can be solved with exact diagonalization. The Hamiltonian of the many-body Hilbert space is constructed completely. Then, the dynamics of the many-body Hamiltonian is calculated by propagating the Schr/"odinger equation ${\ket{\Psi(t)}=\expU{-i\mathcal{H}t}\ket{\Psi(0)}}$ in time, starting with an initial state $\ket{\Psi(0)}$. This method is limited by the Hilbert space size, which increases exponentially with the number of lattice sites and atoms.

{\bfseries DMRG}
The wavefunction of gapped one-dimensional Hamiltonians can be efficiently represented by Matrix product states (MPS). The Hamiltonian is represented as Matrix product operator (MPO). The ground state of the MPS is found by applying the MPO locally on each site, sweeping several times across the different sites of the system. 

The wavefunction is propagated in time by repeated application of $\expU{-i\mathcal{H}\Delta t}$ on the MPS with small time steps $\Delta t$. For non-equilibrium systems this method becomes numerically demanding for larger times, as the entanglement of the wavefunction increases in a strongly excited system. Then, the necessary bond dimension to achieve sufficient accuracy increases over time as well, limiting the maximal time the MPS can be propagated in reasonable computational time. We use the ITensor library to simulate our system\cite{itensor}.
\revC{In our specific setup, we model the source and drain leads as extended one-dimensional hard-core Bose-Hubbard chains attached to the ring. The source is given by
	\begin{equation*}\label{HamiltonSource}
	\mathcal{H}_\text{S}^\text{DMRG}=-\sum_{j=0}^{L_\text{S}-1}\left(J\cn{s}{j}\an{s}{j+1} + \text{H.C.}\right)+\sum_{j=0}^{L_\text{S}-1}\frac{U}{2}\nn{s}{j}(\nn{s}{j}-1)\;,
	\end{equation*}
	where $\an{s}{j}$  and $\cn{s}{j}$  are the annihilation and creation operator at site $j$ in the source leads, ${\nn{s}{j}=\cn{s}{j}\an{s}{j}}$ is the particle number operator, $J$ is the intra-lead hopping, $L_\text{S}$ the number of source lead sites and $U$ is the on-site interaction between particles. The drain lead has a similar Hamiltonian, with length $L_\text{D}$ and respective operators $\an{d}{j}$ and $\cn{d}{j}$. 
	The coupling Hamiltonian between the source lead and ring, and ring and drain lead is
	\begin{equation*}
	\mathcal{H}_\text{I}^\text{DMRG}=-K\cn{s}{0}\an{a}{0}-K\cn{d}{0}\an{a}{L/2}+ \text{H.C.}\;,
	\end{equation*}
	where $K$ is the coupling strength. The ring Hamiltonian is the same as defined in the main text.}

\sectionappendix{Analytic results on the time-dynamics of non-interacting particles}\label{NACalculation}
In this section, we derive analytic results for the time evolution in the dynamics between source and drain. 
To investigate the Hamiltonian, it is convenient to write the ring Hamiltonian in Fourier space (for ${U=0}$)
\begin{align*}
\mathcal{H}_\text{r}={}&\sum_{j=0}^{L-1}-2J\cos{\left(\frac{2\pi}{L}(j-\Phi)\right)}\cn{b}{j}\an{b}{j}\\
{}&+\sum_{j=0}^{L-1}\frac{K}{\sqrt{L}}\left(\cn{a}{\text{S}}\an{b}{j}+(-1)^j\cn{a}{\text{D}}\an{b}{j}+\text{H.C.}\right)\;,\numberthis\label{HMode}
\end{align*}
where $\an{b}{j}$ ($\cn{b}{j}$) the Fourier transform of the annihilation (creation) operator of the ring. 

In the following, we assume that all the particles are initially loaded into the source.
We investigate the dynamics for weak coupling ${K\ll J}$ or for small number of ring sites $L$. 
The time evolution is governed by the eigenmodes $\ket{\Psi_\text{j}}$ with energy $E_j$. It is given by ${\ket{\Psi(t)}=\sum_j\expU{-iE_jt}\ket{\Psi_\text{j}}\bra{\Psi_\text{j}}\ket{\Psi(0)}}$. We define the coefficient of the overlap between eigenstate and initial condition ${A_j=\braket{\Psi_\text{j}}{\Psi(0)}}$. Due to the symmetry of the system, the spectrum has pairs of eigenvalues $\pm E_j$. We find that the absolute value of $A_j$ is the same for these pairs (the sign depends on the parity), in both source and drain. Using these properties, we can write the dynamics for the density in source $n_\text{S}$ for each eigenvalue pair, where $j$ is summed over $L/2+1$ eigenvalue pairs in ascending order of the eigenvalues
\begin{equation*}
n_\text{S}(t)=\abs{\sum_{j\in\text{pairs}} \cos(E_jt)\abs{A_j}^2}^2
\end{equation*}
and the drain density $n_\text{D}$ for odd parity
\begin{equation*}
n_\text{D}(t)=\abs{\sum_{j\in\text{pairs}} (-1)^j\sin(E_jt)\abs{A_j}^2}^2
\end{equation*}
and for even parity
\begin{equation*}
n_\text{D}(t)=\abs{\sum_{j\in\text{pairs}} (-1)^j\cos(E_jt)\abs{A_j}^2}^2\;.
\end{equation*}

In the weak coupling limit ${K\ll J}$ or for small number of ring sites $L$, the dynamics can be described by an effective, reduced system, consisting of source, drain and a small number of ring eigenmodes $L_0$. With above equations, the dynamics is then given by the sum over the ${L_0/2+1}$ eigenmode pairs of the reduced system. 

We justify this method as follows:
Initially, all particles are prepared in the source and a total energy ${E=0}$ (we assume that source and drain have zero potential energy).
Source and drain are coupled via the ring eigenmodes. Coupling is most efficient if the energy difference between the modes is small, thus the leads will couple mainly to the ring eigenmodes with energy close to the leads.

We identify two mechanism for transport through the ring: Firstly, resonant coupling to ring eigenmodes with energies close to that of the uncoupled leads. A similar concept is known as resonant tunneling in quantum dots. Secondly, off-resonant coupling enhanced by interference via all ring modes with energies not too close to the leads. 
Which mechanism is important depends on $L$ and can be grouped into two distinct parities ${L/2}$ even and odd. For even parity, off-resonant coupling is not possible as it turns out the ring modes destructively interfere for any value of $\Phi$. Thus, here transport is dominated by resonant coupling to ring eigenmodes ${E\approx0}$. 
For odd parity, both mechanisms contribute. Here, for ${\Phi\approx0}$, off-resonant coupling is dominant as there are no ring modes close to ${E=0}$, while for ${\Phi\approx\frac{1}{2}}$ resonant coupling is dominant as there are two ring modes on resonance at ${E=0}$. The difference in both parities arises from two effects: First, there are different eigenmode distributions. For example, for even parity and ${\Phi=0}$ there are two ring modes with eigenvalue zero, whereas for odd parity this is not the case. For ${\Phi=0.5}$, the opposite is true, e.g. odd parity has two ring modes with zero energy, whereas even parity has not.

The second contribution is the interference of ring modes at the ring-drain coupling. The eigenvalues can be grouped into pairs with the same absolute value, but opposite sign $\pm E$. They have the momentum mode number $n_-=n$ and $n_+=n+\frac{L}{2}$. As seen in Eq.\ref{HMode}, the sign of the ring-drain coupling depends on the momentum mode number. For even parity, an eigenvalue pair $\pm E$ has the same sign for the ring-drain coupling, whereas for odd the coupling for the two eigenvalues of the pair has opposite signs. This has a profound effect on the dynamics. For even parity, the eigenmode pair $\pm E$ will interfere destructively in the drain, while for odd constructively. This effect is independent of $\Phi$. Note that when the ring modes couple strongly with the leads, the interference condition is relaxed (as the ring modes hybridize with source and drain). Thus, this argument is only strictly valid for off-resonant coupling.
Both descriptions break down in the strong-coupling limit and for a large number of sites as more and more ring eigenmodes couple to the leads as the energy spacing between ring modes decreases.

Next, we describe how to calculate the eigenvalues of ring-lead system.
We write down the eigenvalue equation with the Hamiltonian of Eq.\ref{HMode} for noninteracting particles ${\op{H}\ket{\Psi}=E\ket{\Psi}}$ for an arbitrary eigenstate $\ket{\Psi}=\alpha_\text{S}\ket{\text{S}}+\alpha_\text{D}\ket{\text{D}}+\sum_{n=0}^{L-1}\alpha_n\ket{n}$, where $\alpha_\text{S}$ is the coefficient of the source site, $\alpha_\text{D}$ of the drain, and $\alpha_\text{n}$ of ring eigenmode $n$. We get ${L+2}$ equations for the coefficients
\begin{align*}
E\alpha_\text{S}={}&\frac{K}{\sqrt{L}}\sum_{n=0}^{L-1}\alpha_n\\
E\alpha_\text{D}={}&\frac{K}{\sqrt{L}}\sum_{n=0}^{L-1}(-1)^n\alpha_n\\
E\alpha_n={}&-2J\cos\left(\frac{2\pi (n-\Phi)}{L}\right)\alpha_n+\frac{K}{\sqrt{L}}(\alpha_\text{S}+(-1)^n\alpha_\text{D})\;.
\end{align*}
We insert the equations for $a_\text{n}$ into the source and drain equations
\begin{align*}
E\alpha_\text{S}={}&\frac{K^2}{2JL}\sum_{n=0}^{L-1}\frac{\alpha_\text{S}+(-1)^n\alpha_\text{D}}{\frac{E}{2J}+\cos\left(\frac{2\pi (n-\Phi)}{L}\right)}\\
E\alpha_\text{D}={}&\frac{K^2}{2JL}\sum_{n=0}^{L-1}\frac{(-1)^n\alpha_\text{S}+\alpha_\text{D}}{\frac{E}{2J}+\cos\left(\frac{2\pi (n-\Phi)}{L}\right)}\;.\numberthis\label{EVEquation}
\end{align*}
Using these two equations as starting point, we can make approximations. We define ${\tilde{K}=K/J}$.

For resonant coupling, we keep only the dominant terms of the sum in Eq.\ref{EVEquation}. E.g. for odd parity and ${\Phi\approx0.5}$ they are ${n=\frac{L+2}{4}}$ and ${n=\frac{3L+2}{4}}$, and for even parity with ${\Phi\approx0}$  ${n=\frac{L}{4}}$ and ${n=\frac{3L}{4}}$. The resulting eigenvalues for even parity are
\begin{equation}
E=0,\;\;E/J=2\sqrt{\sin\left(\frac{2\pi\Phi}{L}\right)^2+\left(\frac{\tilde{K}}{\sqrt{L}}\right)^2}\;,\label{Freq4nApp}
\end{equation}
and for odd
\begin{equation}
E_\pm/J=\sin\left(\frac{2\pi(\Phi-0.5)}{L}\right)\pm\sqrt{\sin\left(\frac{2\pi(\Phi-0.5)}{L}\right)^2+\left(\frac{\sqrt{2}\tilde{K}}{\sqrt{L}}\right)^2}\;.
\end{equation}
The equations are quite different for each parity as the sign of the ring-lead coupling depends on the parity as well.

For even parity, $\Phi\approx0.5$ and weak coupling the source oscillations disappear in the weak-coupling limit. Interestingly, for even parity and strong coupling we still find density oscillations and we observe a characteristic beating. To calculate, we have to include in total four wavenumbers ($n=\frac{L}{4}$, $n=\frac{L+4}{4}$, and $n=\frac{3L}{4}$, $n=\frac{3L+4}{4}$) as at this flux all these ring modes have nearly the same energy. We get the same eigenvalues as in Eq.\ref{Freq4nApp} and additionally
\begin{equation}
E/J=2\sqrt{\sin\left(\frac{2\pi(\Phi-1)}{L}\right)^2+\left(\frac{\tilde{K}}{\sqrt{L}}\right)^2}\;.\label{Freq4nAdd}
\end{equation}
The beating frequency is given by the subtraction of these eigenvalues and Eq.\ref{Freq4nApp}.

In the other limit $\Phi\approx0$ and $K/J\approx1$ or $L\gg1$, a higher frequency mode appears in the density oscillation. To calculate, we include in total six wavenumbers ($n=\frac{L}{4}$, $n=\frac{L+4}{4}$, $n=\frac{L-4}{4}$, and $n=\frac{3L}{4}$, $n=\frac{3L+4}{4}$, $n=\frac{3L-4}{4}$). We get the original eigenvalue of Eq.\ref{Freq4nApp} and two new ones, of which only the following has a non-negligible coefficient $A_j$
\begin{align*}
E/J={}&\left[\left(\frac{2\tilde{K}}{\sqrt{L}}\right)^2+2\left(1-\cos\left(\frac{4\pi}{L}\right)\cos\left(\frac{4\pi\Phi}{L}\right)\right)\right.\\
&+\left\{\left(\frac{2\tilde{K}}{\sqrt{L}}\right)^4+\left(1+\cos\left(\frac{8\pi}{L}\right)\cos\left(\frac{8\pi\Phi}{L}\right)\right.\right.\\
&\left.\left.\left.-\cos\left(\frac{8\pi}{L}\right)-\cos\left(\frac{8\pi\Phi}{L}\right)\right)\right\}^{\frac{1}{2}}\right]^{\frac{1}{2}}\;.\numberthis\label{Freq4nHigher}
\end{align*}
This equation is not valid for small $L$ and $\Phi\approx0.5$.

Next, we show how to calculate off-resonant contributions. These are only present for odd parity, and do not play a role for even parity.
In the weak-coupling limit, we assume that the eigenenergy $E$ of the full system will be close to the energy of the uncoupled leads ${E=0}$. This assumption is valid as the ring modes are far detuned from the leads. Thus, we perform a Taylor expansion of the fraction around ${E=0}$
\begin{equation*}
\frac{1}{\frac{E}{2J}+\cos\left(\frac{2\pi (n-\Phi)}{L}\right)}=\sum_{p=0}^{\infty}\left(-\frac{E}{2J}\right)^p\sec^{p+1}\left(\frac{2\pi (n-\Phi)}{L}\right)\;.
\end{equation*}
The eigenvalue equation becomes
\begin{equation*}
E\alpha_\text{S}=\frac{K^2}{2JL}\sum_{p=0}^{\infty}\left(\frac{-E}{2J}\right)^p(\beta_p^+\alpha_\text{S}+\beta_p^-\alpha_\text{D})
\end{equation*}
and analog for $E\alpha_\text{D}$. The symmetric and anti-symmetric combination of both equations give
\begin{equation*}
E_\pm=\frac{K^2}{2JL}\sum_{p=0}^{\infty}\left(\frac{-E}{2J}\right)^p(\beta_p^+\pm\beta_p^-)
\end{equation*}
We define
\begin{equation*}
\beta_p^+=\sum_{n=0}^{L-1}\sec(\frac{2\pi}{L}(n-\Phi))^{p+1}
\end{equation*}
and 
\begin{equation*}
\beta_p^-=\sum_{n=0}^{L-1}(-1)^n\sec(\frac{2\pi}{L}(n-\Phi))^{p+1}\;.
\end{equation*}
The coefficients reveal the parity effect in $L/2$. For even parity and all values of $\Phi$, $\beta_p^+$ and $\beta_p^-$ is zero or infinity for  ${p\in\text{even}}$, which suppresses the oscillations.
For the case ${p\in\text{odd}}$, $\beta_p^+$ and $\beta_p^-$ have nearly the same absolute value. Here, we find that there is always an eigenvalue ${E=0}$. The corresponding oscillation period between source and drain for this energy is $T\rightarrow\infty$. Thus, for even parity in $L/2$ off-resonant coupling can be neglected for any order of $p$ .
For odd parity in $L/2$, we have ${\beta_p^+=0}$ for ${p\in\text{even}}$ and ${\beta_p^-=0}$ for ${p\in\text{odd}}$, else the coefficients are non-zero. Two exact solutions are known to us $\beta_0^-=\pm\frac{L}{\cos(\Phi\pi)}$ and $\beta_1^+=\frac{L^2}{2\cos(\Phi\pi)^2}$.
For a simple zero order expansion of all modes, without resonant tunneling (valid for $\Phi\approx0$), the energy difference of symmetric and anti-symmetric mode and the oscillation frequency between drain and source is
\begin{equation}
E/J=\frac{\tilde{K}^2}{\abs{\cos(\Phi\pi)}}\;.\label{FreqLowFlux4n2}
\end{equation}
It is possible to also derive higher order versions of this equation for increased accuracy. First order yields
\begin{equation}
E/J=\frac{4\tilde{K}^2\cos(\Phi\pi)}{8\cos(\Phi\pi)^2+2L\tilde{K}^2}~.
\end{equation}

It is possible to combine resonant and off-resonant coupling. One has to apply the method for resonant coupling, and use the off-resonant method for all other eigenmodes minus the resonant ones. The sum over $n$ has to be adjusted accordingly. The result for order ${p=0}$ is 

\begin{align*}
&E_\pm/J=\pm\left[\frac{\tilde{K}^2}{2L\sin(\frac{\pi}{L}(1-2\Phi))}-\frac{\tilde{K}^2}{4\cos(\Phi\pi)}+\sin\left(\frac{\pi}{L}(1-2\Phi)\right)\right]\\
&+\left[\left(\frac{\tilde{K}^2}{2L\sin(\frac{\pi}{L}(1-2\Phi))}-\frac{\tilde{K}^2}{4\cos(\Phi\pi)}+\sin\left(\frac{\pi}{L}(1-2\Phi)\right)\right)^2\right.\\
&\left.+\frac{\tilde{K}^2\sin(\frac{\pi}{L}(1-2\Phi))}{\cos(\pi\Phi)}\right]^{\frac{1}{2}}\;.\numberthis\label{FreqFull4n2}
\end{align*}
and up to order ${p=1}$ we get with ${\delta=\csc\left(\frac{\pi}{L}(1-2\Phi)\right)}$ and ${\gamma=\sec(\Phi\pi)}$
\begin{align*}
&E_\pm/J=\pm\frac{L}{\delta}\frac{8+\tilde{K}^2 L\gamma^2-2\tilde{K}^2\delta\gamma}{4\tilde{K}^2\delta^2-L(8+\tilde{K}^2L\gamma^2)}+\numberthis\label{FreqFull4n2Higher}\\
&\left[\left(\frac{L}{\delta}\frac{8+\tilde{K}^2L\gamma^2-2\tilde{K}^2\delta\gamma}{4\tilde{K}^2\delta^2-L(8+\tilde{K}^2L\gamma^2)}   \right)^2+\frac{L}{\delta}\frac{8\tilde{K}^2 \gamma}{8L+\tilde{K}^2(L^2\gamma^2-4\delta^2)}\right]^{\frac{1}{2}}\;.
\end{align*}
Eq.\ref{FreqFull4n2Higher} describes the oscillation frequencies between source and drain for odd parity accurately over a wide parameter range.

So far, we only discussed the eigenvalues, which represent the frequency of the oscillation in source and drain. As outlined in the beginning of the section it can be described by a superposition of Cosines with the relevant eigenvalues of the reduced system. Now, we discuss the relative strength of the Cosine contribution $A_\pm$, which is the coefficient of the eigenvector in drain and source. For resonant coupling, we can write down the reduced Hamiltonian and solve for the eigenvectors and calculate $A_j$. For off-resonant coupling, writing down the reduced Hamiltonian is not so trivial. Using numerics, we established that for odd parity and $\Phi\approx0$, $E_+$ dominates and only a minor contribution of $E_-$ contributes for larger ring lengths. For $\Phi\approx0.5$, both frequencies contribute equally, and a beating between the two frequencies occurs.

\begin{figure}[htbp]
	\centering
	\subfigimg[width=0.24\textwidth]{a}{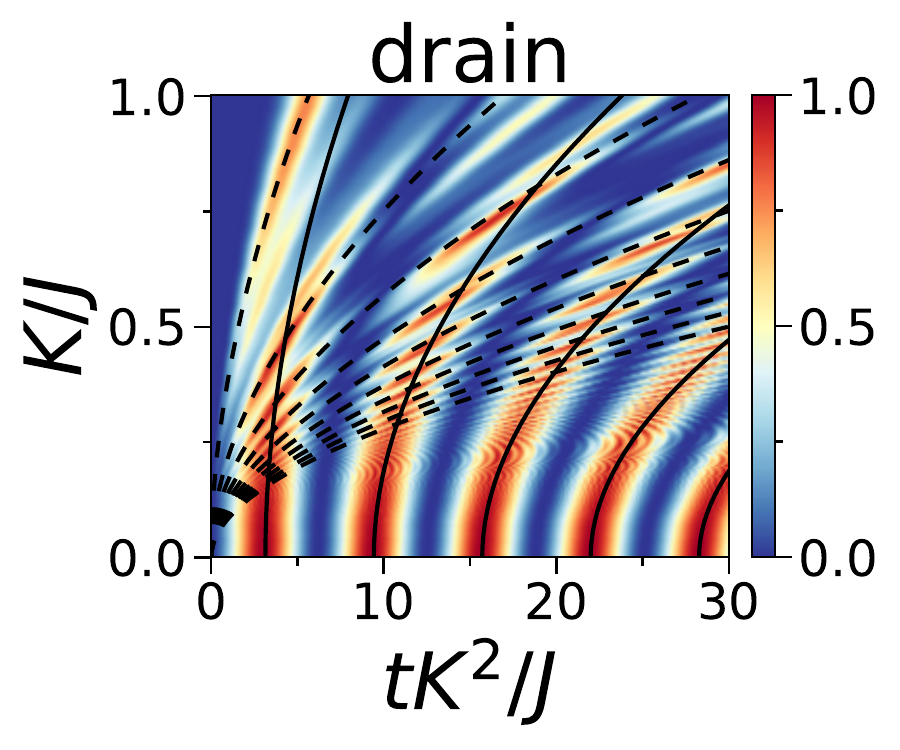}\hfill
	\subfigimg[width=0.24\textwidth]{b}{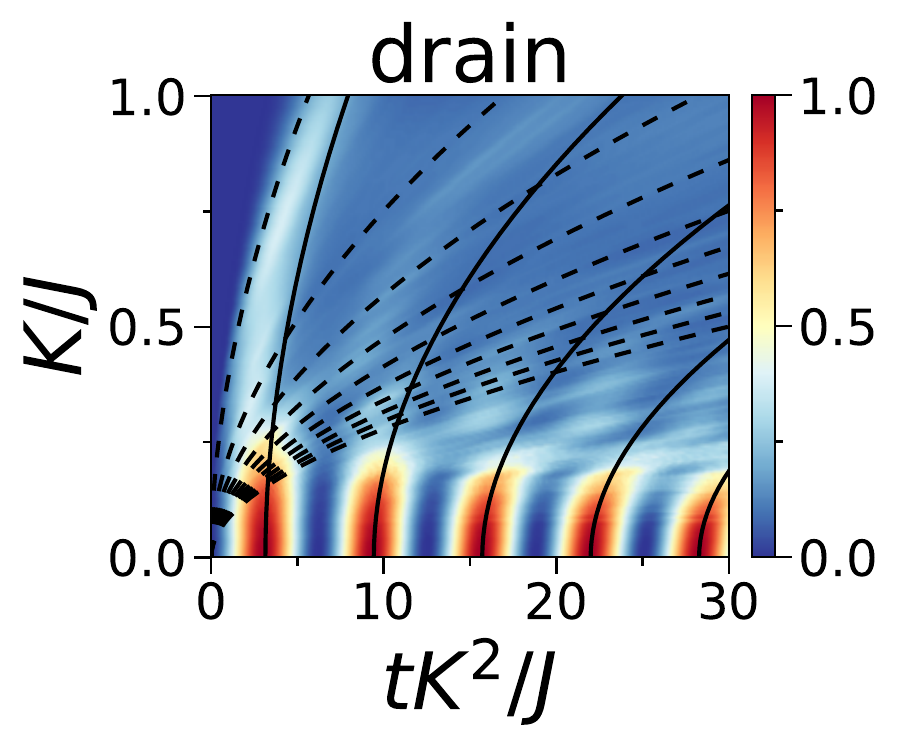}
	\caption{Density in drain against $K/J$ for ${L/2=7}$ and four particles with \idg{a} ${U=0}$ and \idg{b} ${U/J=5}$ with analytic solution using Eq.\ref{FreqFull4n2Higher}.  For weak ring-lead coupling only one oscillation mode contributes, whereas for strong coupling additional oscillation modes are excited.}
	\label{UPlots4n2}
\end{figure}

\sectionappendix{Comparison numerics and analytics}
We plot the ring-lead coupling $K$ for different $U/J$ in Fig.\ref{UPlots4n2}. For weak-coupling $K/J$, we observe source drain oscillation which are nearly independent of $U/J$, as the ring population in the weak-coupling limit is low, suppressing particle-particle scattering. With increasing $K/J$, we notice faster oscillation patterns. They smear out with increasing interaction as the ring fills up for stronger ring-lead coupling.

For two atoms, the ring spectrum can be derived analytically \cite{valiente2008two} in terms of scattering (real valued relative quasi-momentum $k$) and bound states (imaginary $k$). The bound states have only half the flux quantum compared to single atoms (like Cooper-pairs) and therefore interfere constructively when propagating through the ring at the degeneracy point. For very weak interaction, the oscillations in a simplified model of a ring without leads can be calculated from the energy difference of bound and scattering states at the edge of the Brillouin zone. We find that the analytic results matches the main oscillation period of the full numerical calculation.

\sectionappendix{Two potential barriers} 
\rev{We show supplemental data for the first application: Inserting two potential barriers in the ring. Here, we study this application without interaction. By introducing two potential barriers $\Delta$ symmetrically in the center of upper and lower arm of the ring, we can mimic the effect of flux on the oscillation periods. }

\begin{figure}[htbp]
	\centering
	\subfigimg[width=0.24\textwidth]{a}{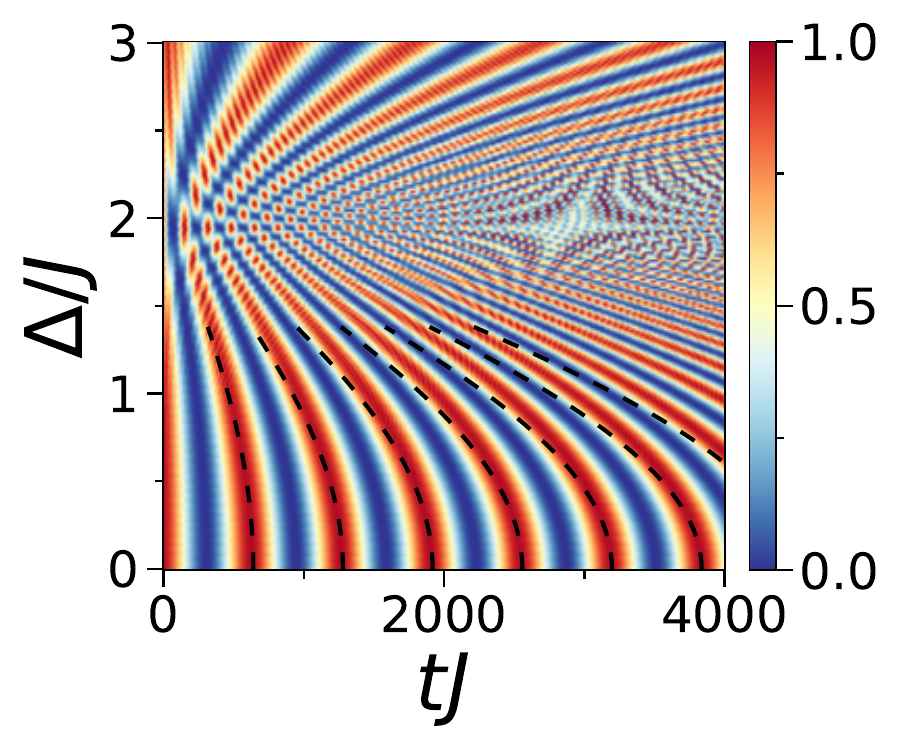}\hfill
	\subfigimg[width=0.24\textwidth]{b}{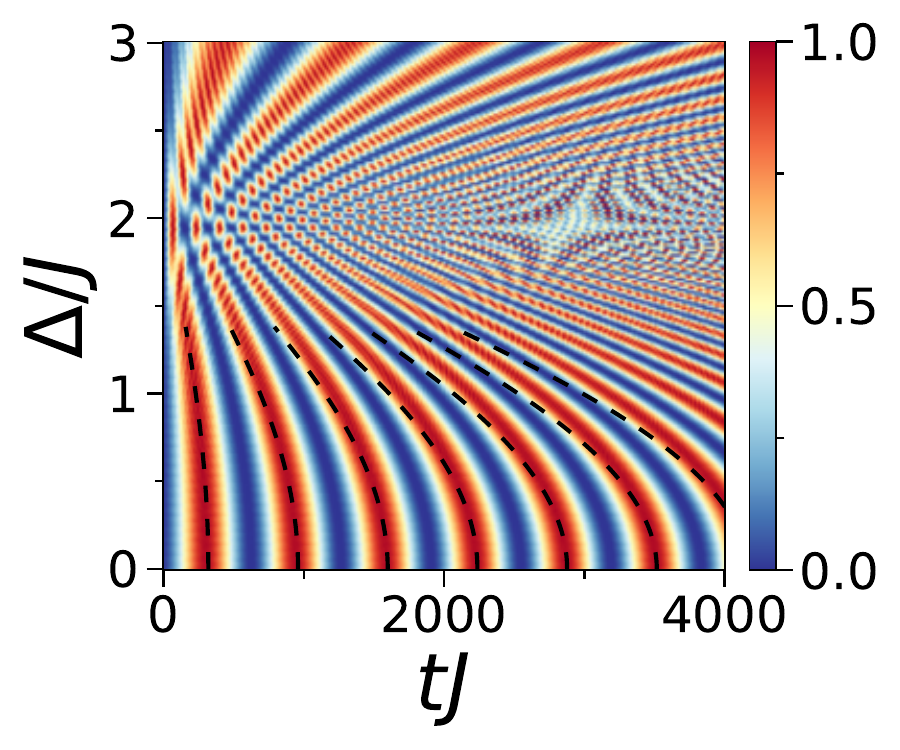}
	\caption{Density in source and drain against two potential barriers in each arm of ring with potential depth $\Delta/J$ for \idg{a,b} ${L/2=7}$ without interaction ${U=0}$. Other parameter are ${K/J=0.1}$, ${\Phi=0}$. The specific placement of the barriers has a strong impact on the dynamics. We choose a special placement which produced a behaviour which mimics the flux. The fit uses the formulas for flux, but replaces $\Phi=\Delta/(2J)$. 
		The barriers behave like a scattering impurity, which couple different momentum modes. $\Delta$ mainly changes the frequency of the density oscillation, but does not add any new additional oscillation frequencies. This means that the system is still dominated by one eigenmode. The scattering is only perturbing the overall dynamics, but not adding higher frequencies (in contrast to increasing ring-lead coupling $K$ or adding more ring sites).}
	\label{DeltaPlotsNA}
\end{figure}

We plot the density in source and drain against $\Delta$ in Fig.\ref{DeltaPlotsNA} without interaction ${U=0}$. The oscillation frequency follows nearly the same relation as for flux, but no destructive interference is observed. This suggest we are not in the tunneling regime, and potential barriers and weak link play the role of scatters, and do not induce a simple phase shift.

\begin{figure}[htbp]
	\centering
	\subfigimg[width=0.24\textwidth]{a}{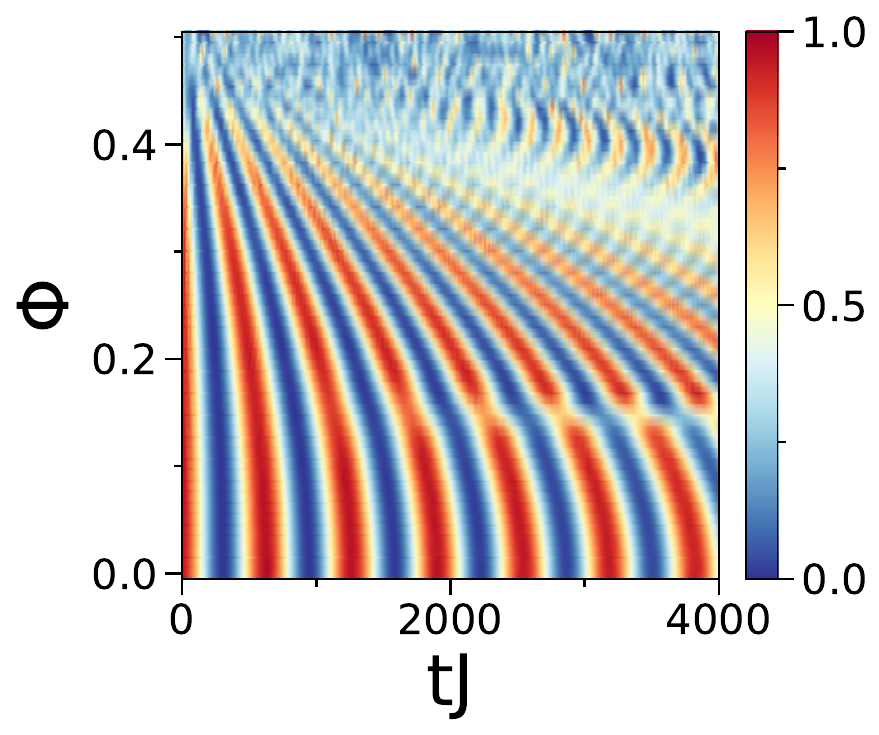}\hfill
	\subfigimg[width=0.24\textwidth]{b}{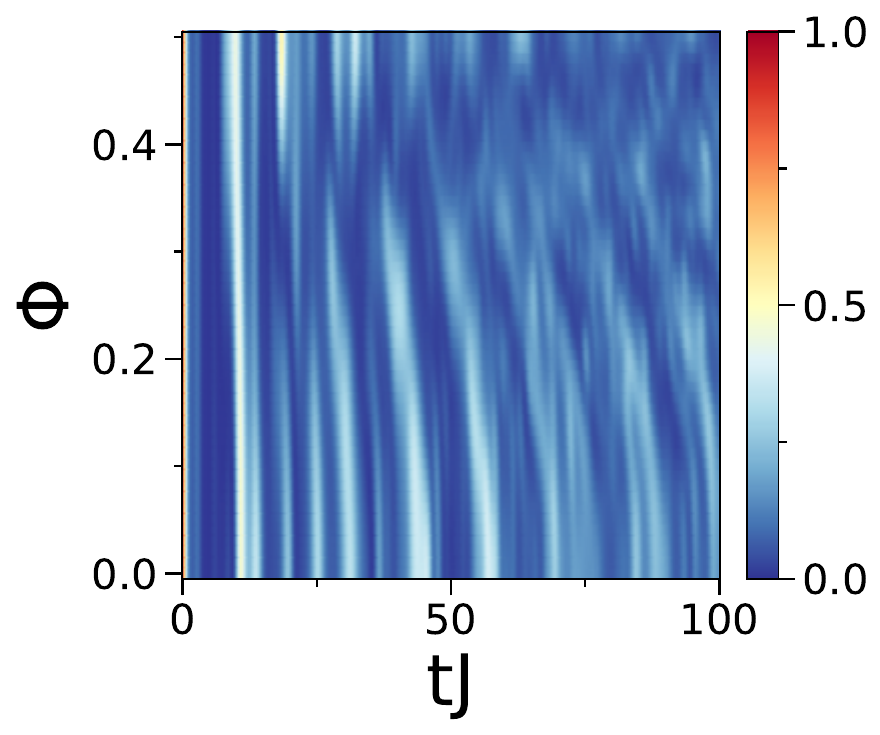}\\
	\subfigimg[width=0.24\textwidth]{c}{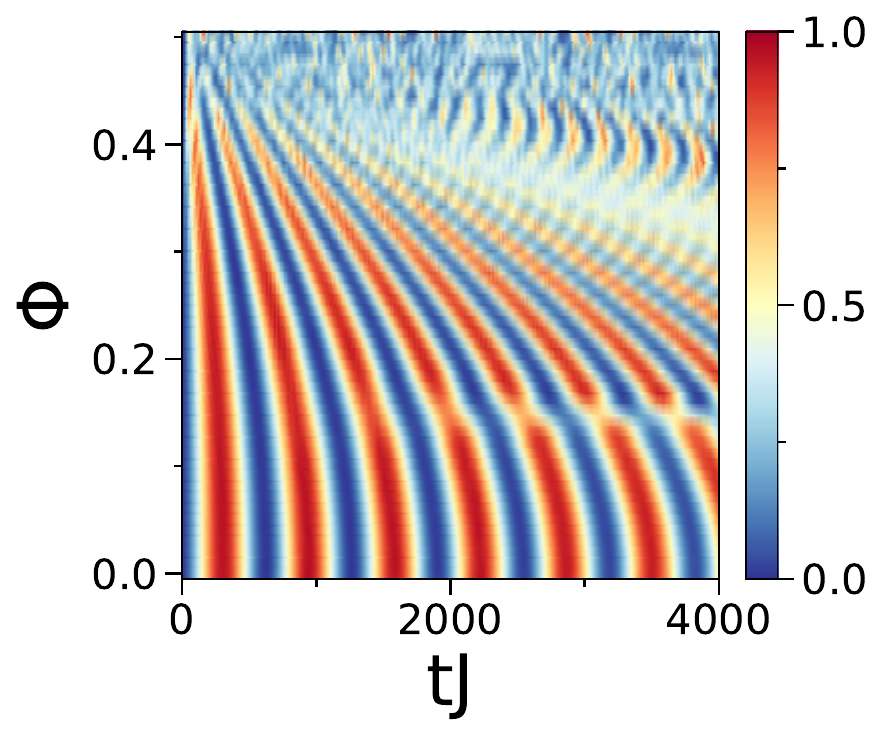}\hfill
	\subfigimg[width=0.24\textwidth]{d}{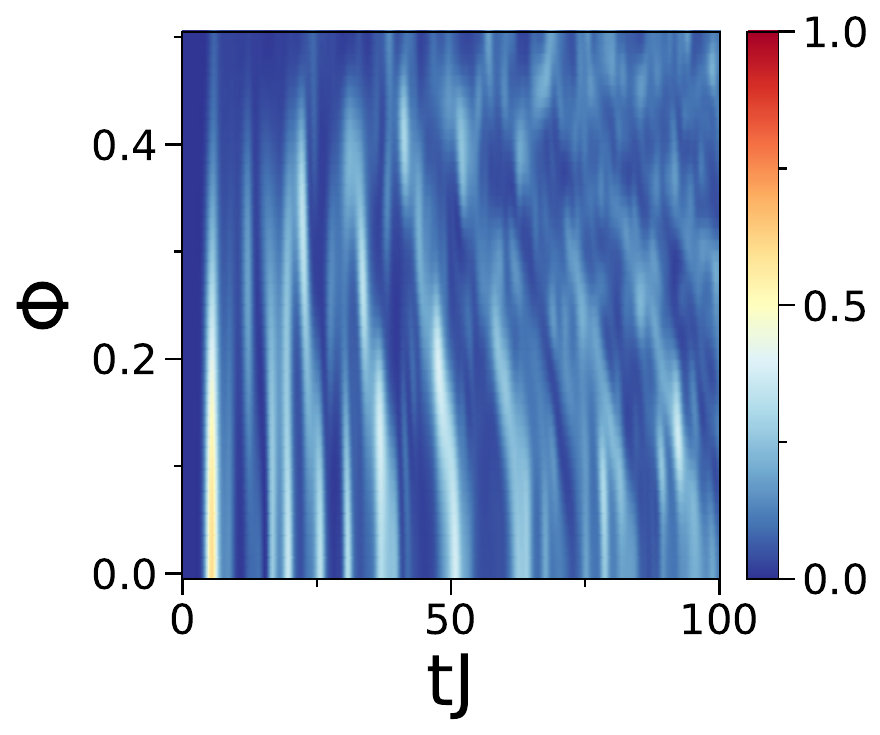}\\
	\subfigimg[width=0.24\textwidth]{e}{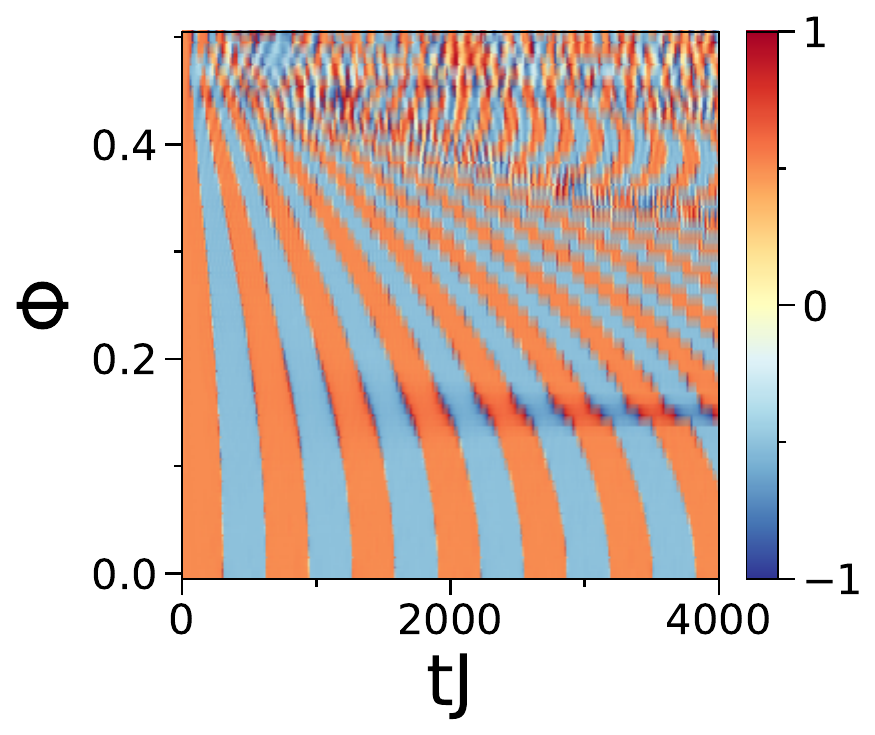}\hfill
	\subfigimg[width=0.24\textwidth]{f}{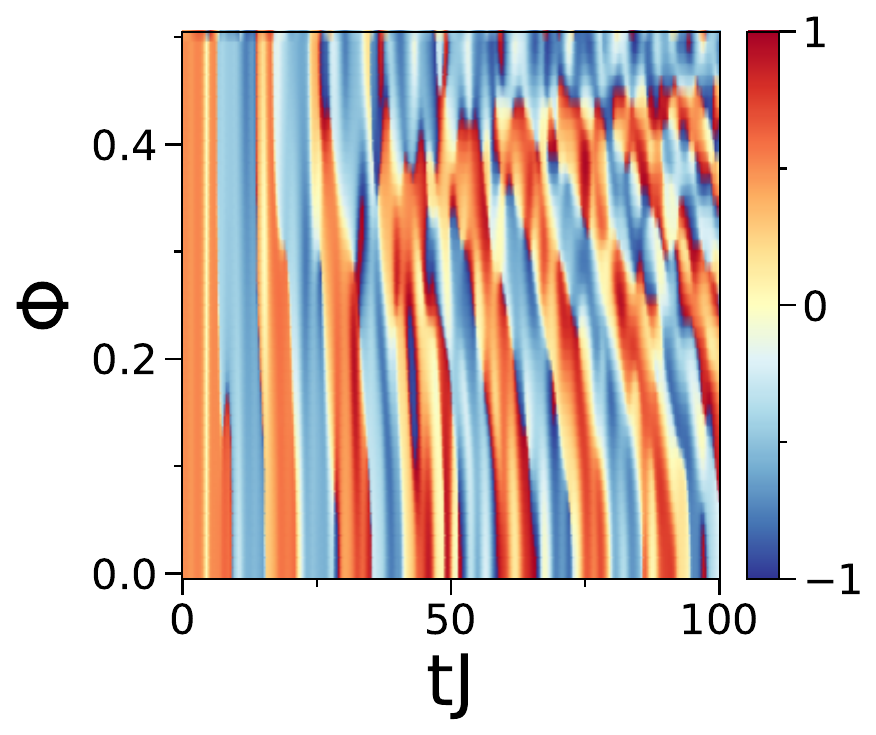}
	\caption{Time evolution of ring-lead system. Density of source \idg{a,b}, drain \idg{c,d} and relative phase of source and drain $\Delta \phi/\pi$ \idg{e,f} plotted against flux $\Phi$. \idg{a,c,e} weak ring-lead coupling ${K/J=0.1}$ (on-site interaction ${U/J=5}$). \idg{b,d,f} strong ring-lead coupling ${K/J=1}$ (${U/J=0.5}$). Time is indicated $tJ$ in units of inter-ring tunneling parameter $J$. The number of ring sites is ${L=14}$ with $N_\text{p}=2$ particles initially prepared in the source. }
	\label{phase}
\end{figure}

\sectionappendix{Phase dynamics}
In this section, we plot the phase dynamics in the ring-lead system. 
The source is initialized with a specific number of particles, and thus the phase is initially not well defined. During the evolution, the number of particles at each site becomes uncertain to some degree, and we can define a phase via the two-body correlators $\cn{a}{n}\an{a}{m}$. The operators can be mapped to complex numbers $\cn{a}{n}\sim\sqrt{\avg{\cn{a}{n}\an{a}{n}}}\expU{i\phi_n}$ with phase $\phi_n$. We find $\avg{\cn{a}{n}\an{a}{m}}\sim\sqrt{\avg{\cn{a}{n}\an{a}{n}}\avg{\cn{a}{m}\an{a}{m}}}\expU{i(\phi_n-\phi_m)}$. So the phase of the two-body correlator is a direct measure of the relative phase between two sites. In Fig.\ref{phase}, we plot the phase between source and drain $\Delta \phi=\phi_\text{source}-\phi_\text{drain}$. We observe that the phase dynamics is very similar to the density dynamics for weak coupling and also to lesser degree for strong coupling.

The phase can also be related to the expectation value of the current, e.g. the source-ring current $\avg{j}=-iK(\avg{\cn{a}{\text{S}}\an{a}{0}}-\text{h.c.})\sim2K\sqrt{\avg{\cn{a}{S}\an{a}{S}}\avg{\cn{a}{0}\an{a}{0}}}\sin(\phi_S-\phi_0)$. Plots of the current is shown in the main text.

\end{document}